# SUPER MARX GENERATOR

# FOR THERMONUCLEAR IGNITION

F. Winterberg

University of Nevada, Reno

August 2008

**Abstract**

In ongoing electric pulse power driven inertial confinement fusion experiments, Marx generators are connected in parallel with the target in the center of a ring of the Marx generators. There the currents, not the voltages add up. Instead of connecting a bank of Marx generator in parallel, one may connect them in series, adding up their voltages, not the currents. If, for example, fifty 20 MV Marx generators are connected in series, they would add up to a gigavolt. But to prevent breakdown, the adding up of the voltages in such a super-Marx generator must be fast. For this reason, it is proposed that each of the Marx generators charges up a fast discharge capacitor, with the thusly charged fast capacitors becoming the elements of a second stage super Marx generator. In a super Marx generator, the Marx generators also assume the role of the resistors in the original Marx circuit.

With a voltage of $10^9$ Volt and a discharge current of $10^7$ Ampere, the generation of a $10^{16}$ Watt GeV proton beam becomes possible, which can ignite in liquid deuterium a thermonuclear micro-detonation.

# 1. Introduction

In an inertial confinement fusion a high gain is required. This poses a problem for laser fusion, where the photon flash from the thermonuclear micro-explosion is likely to destroy the optical laser ignition apparatus. Heavy ion beam ignition fares better, but it requires very large conventional particle accelerators, apart from the problem to stop the ion beam over a short distance, both in the "direct" and "indirect" (black body radiation induced) target implosion drive. Using a bank of Marx generators surrounding a DT target requires replaceable transmission lines, an unappealing feature, but the attainment of 30 MV would make possible to replace laser beams with light ion beams [1]. This would eliminate the need for replaceable transmission lines, with the beams extracted from magnetically insulated diodes driven by Marx generators [2]. The reason why high voltages are here important is that a beam gets stiffer, as the voltage driving the beam gets higher. A stiffer beam implies that the distance of the diode from the target can be made larger, of importance for a repetition rate operation. This illustrates the importance to aim for high voltages. Going to even higher voltages would still be better.

A voltage of $10^9$ Volt would be ideal, because there a pulsed $10^7$ Ampere GeV proton beam could be generated, with the current large enough to entrap in the target the fusion reaction alpha particles by the magnetic field of the beam. Gigajoule energies, which at these voltages can be released in less than $10^{-7}$ seconds, at a power of $10^{16}$ Watt, would make possible the ignition of a pure DD thermonuclear reaction, without any tritium. One gigajoule is on a logarithmic scale in the middle in between the one megajoule energy required for the ignition of a DT reaction, and the $10^3$ gigajoule (fission bomb) energy used for the ignition of a DD reaction (Mike test). For an ignition energy of one gigajoule and a gain of ~ $10^2$, an energy of 100 gigajoule =$10^{18}$ erg would be released. Such a large yield would require to confine the explosion in a cavity with a radius of more than 10 meters. If the target is in the center of the cavity, the length of the beam reaching the target must be comparable to the diameter of the cavity. To direct the beam over such a distance onto a target, the beam must be stiff. This requires very high voltages.

In the DT reaction 80% of the energy goes into neutrons. In the DD reaction the proportion is much smaller. The DT reaction depends on tritium, which must be obtained by the breeding of lithium, a relatively rare element. Deuterium is abundant and everywhere available. All this suggests that the future of fusion is in the ignition and burn of deuterium and by implication in the attainment of very high voltages.

**2. The importance of High Voltages for inertial Confinement Fusion**

The reaching out for high voltages in the quest for the ignition of thermonuclear micro-explosions by inertial confinement can be explained as follows:

1. The energy $e$ [erg] stored in a capacitor $C$ [cm] charged to the voltage $V$ [esu] is equal to

$$e = (1/2)CV^2, \qquad (1)$$

with an energy density

$$\varepsilon \sim e/C^3 \sim V^2/C^2 \qquad (2)$$

The energy $e$ is discharged in the time $\tau$ [sec] ($c$ velocity of light)

$$\tau \sim C/c \qquad (3)$$

with the power $P$ [erg/s]

$$P \sim e/\tau \sim cV^2 \qquad (4)$$

This shows that for a given dimension of the capacitor measured in its length, and hence volume, the energy stored and power released goes in proportion to the square of the voltage.

2. If the energy stored in the capacitor is released into a charged particle beam with the particles moving at the velocity v, the current should be below the critical Alfvén limit.

$$I = \beta\gamma I_A \qquad (5)$$

where $\beta=v/c$, $v$ particle velocity, $\gamma = (1- v^2/c^2)^{-1/2}$ the Lorentz boost factor, and $I_A = mc^3/e$. for electrons $I_A = 17$ kA, but for protons it is 31 MA. For $I << \beta\gamma I_A$, one can view the beam charged particles accompanied carrying along an electromagnetic field, while for $I >> \beta\gamma I_A$ it is better viewed as an electromagnetic pulse carrying along with it some particles. For $I >> \beta\gamma I_A$, the beam can propagate in a space-charge and current-neutralizing plasma, but only if $I \leq \beta\gamma I_A$ can the beam be easily focused onto a small area, needed to reach a high power flux density. If a power of ~ $10^{15}$ Watt shall be reached with a relativistic electron beam produced by a $10^7$ Volt Marx generator, the beam current would have to be $10^8$ Ampere. For 10MeV electrons one has $\gamma \cong 20$ and $\beta\gamma I_A \sim 3\times10^5$ Ampere, hence $I >> \beta\gamma I_A$. But if the potential is $10^9$ Volt, a proton beam accelerated to this voltage and with a current of $I = 10^7$ A is below the Alfvén current limit for protons. It would have the power of $10^{16}$ Watt, sufficiently large to ignite a deuterium thermonuclear reaction.

### 3. Super Marx Generator

Would it be not for breakdown, one could with a Marx generator reach in principle arbitrarily large voltages. According to Paschen's law, the breakdown voltage in gas between two plane parallel conductors is only a function of the product $pd$, where $p$ is the gas pressure and $d$ the distance between the conductors. For dry air at a pressure of 1 atmosphere the breakdown voltage is $3\times10^4$ V/cm, such that for a pressure of 100 atmospheres the breakdown voltage would be $3\times 10^8$ V/cm. For a metersize distance between the conductors this implies a potential difference of the order $10^9$ Volt. But as in lightning, breakdown occurs at much lower voltages by the formation of the "stepped leader". The formation of a stepped leader though requires some time. Therefore, if the buildup of the high voltage is fast enough, breakdown by a stepped leader can be prevented. In a Marx generator the buildup of the voltage is not fast enough to reach gigavolt. It is the idea of the super Marx generator how this might be achieved.

To obtain a short discharge time with a single Marx generator, the Marx generator charges up a fast discharge capacitor, discharging its load in a short time. This suggests using a bank of such fast discharge capacitors as the elements of a Marx generator, each one of them charged up by one Marx generator to a high voltage. One may call such a two stage Marx generator a super Marx generator. If $N$ fast capacitors are charged up by $N$ Marx generators in parallel to the voltage $V$, the closing of the spark gap switches in the super Marx generator adds up their voltages to the voltages $NV$. In the super Marx generator, the Marx generators also serve as the resistors in the original Marx circuit. It is also possible to disconnect the Marx generator from the super Marx after up charge is completed.

It is known, and used in electric power interrupters, that a high pressure gas flow can blow out a high power electric arc. Vice verse, it can be expected that a rapid gas flow can prevent breakdown [3]. Therefore, to further reduce the danger of breakdown, one may bring the breakdown preventing gas into fast motion, just prior to the moment the super Marx generator is fired. However, 6 MV rim fire switches are already in use of the Z Facility of Sandia.

## 4. Ignition of Deuterium

By connecting the high voltage terminal of the super Marx generator to a Blumlein transmission line, a very high voltage pulse with a fast rise time can be generated. At the envisioned very high voltages one can make a controlled breakdown in a gas, or liquid, generating an ion beam below the Alfvén limit. At these high voltages ion beams are favored over electron beams, because electron beams are there above the Alfvén limit. To assure that all the ions have the same charge to mass ratio, the gas or liquid must be hydrogen or deuterium, otherwise the beam will spread out axially, losing its maximum power.

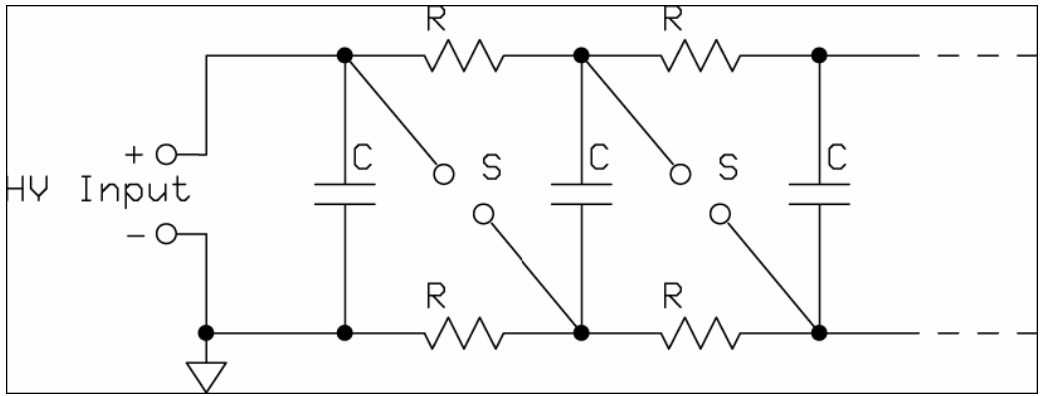

Fig. 1: In an "ordinary" Marx generator n capacitors $C$ charged up to the voltage $v$, and are over spark gaps switched into series, adding up their voltages to the voltage $V = nv$.

Instead of making the breakdown in hydrogen gas, one may let the breakdown happen along a thin liquid hydrogen jet, establishing a bridge between the high voltage terminal of the Blumlein transmission line and the thermonuclear target.

Unlike the traditional spherical implosion type targets for the DT reaction, for the DD reaction cylindrical targets are preferred. Only there is a micro-detonation in pure deuterium possible. This is so because the large electric current of the intense hydrogen beam generates a large azimuthal magnetic field which entraps the charged fusion products within the cylindrical target, the condition for burn and detonation. In a z-pinch, a magnetic field supported detonation along a pinch discharge channel is possible, for pinch currents of the order $10^7$ Ampere [4]. The same is true if the large pinch current of a ~ $10^7$ Ampere is replaced by an ion beam with the same current. For a detonation to occur in deuterium, the burn of the tritium and He³ fusion products of the DD reaction is important [5].

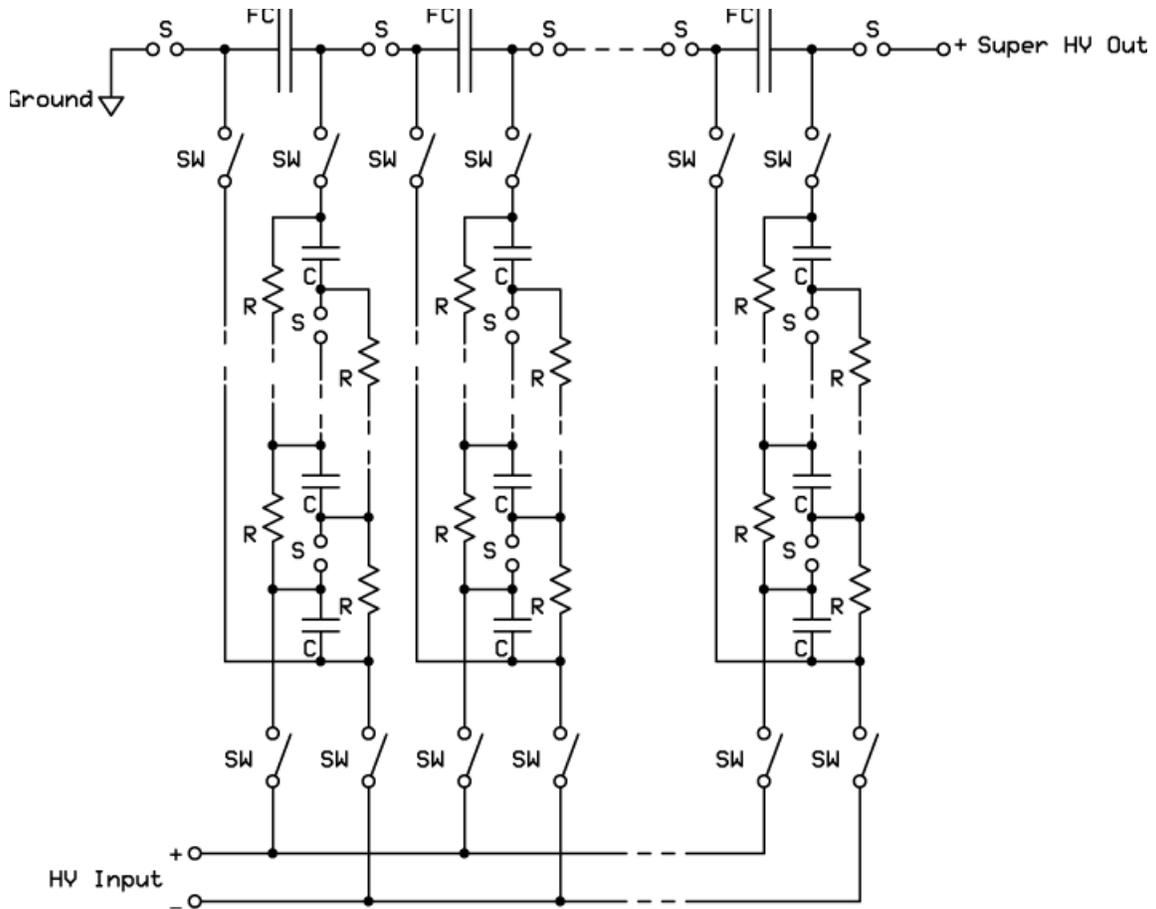

Fig. 2: In a super Marx generator, $N$ Marx generators charge up $N$ fast capacitors FC to the voltage $V$, which switched into series add up their voltages to the voltage $NV$.

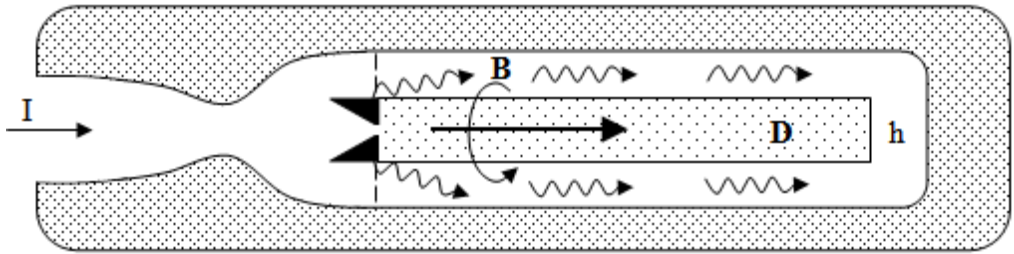

Fig.3: Pure deuterium fusion micro-detonation ignited with an intense proton beam. **D** solid deuterium rod, **h** hohlraum, **I** proton beam, **B** magnetic field.

In a possible configuration shown in Fig.4, the liquid (or solid) D has the shape of a cylinder, placed inside a cylindrical "hohlraum" **h**. A GeV proton beam **I** coming from the left, in entering the hohlraum dissipates part of its energy into a burst of X-rays, compressing the deuterium cylinder, and part of it igniting a detonation wave propagating down the cylinder.

If the rod has the length $z$ and the density $\rho$, the ignition condition for deuterium requires that $\rho z > 10$ g/cm$^2$, at a temperature $T \cong 10^9$ K. Normally, the $\rho z$ condition is given by a $\rho r$ condition for the radius r of a deuterium sphere. Here however, the radial entrapment of the charged DD fusion reaction products, ensured by the magnetic field of the proton beam current in excess of $10^7$ Ampere, replaces the $\rho r > 10$ g/cm$^2$ condition, to a $\rho z > 10$ g/cm$^2$ condition, which is much easier to achieve. The yield of the deuterium explosion then only depends on the total mass of the deuterium.[1]

With both the beam and the target (initially) at a low temperature, the stopping length is determined by the electrostatic two stream instability [6]. In the presence of the strong azimuthal magnetic field, it is enhanced by the formation of a collisionless shock [7]. The stopping range of the protons by the two stream instability is given by

$$\lambda \cong \frac{1.4c}{\varepsilon^{1/3}\omega_i} \qquad (6)$$

where $c$ is here is the velocity of light, $\omega_i$ the proton ion plasma frequency. Furthermore $\varepsilon = n_b/n$, where $n_b$ is the proton number density in the proton beam, and n the deuterium target number density. If the cross section of the beam is 0.1 cm$^2$, one obtains for a $10^7$ Ampere beam that $n_b = 2\times10^{16}$ cm$^{-3}$. For a 100 fold compressed deuterium rod one has $n = 5\times10^{24}$ cm$^{-3}$ with $\omega_i = 2\times10^{15}$ s$^{-1}$. One there finds that $\varepsilon = 4\times10^{-9}$ and $\lambda \cong 1.2\times10^{-2}$ cm. At a deuterium number density $n = 5\times10^{24}$ cm$^{-3}$, the deuterium density is $\rho = 17$ g/cm$^3$. To have $\rho z \geq 10$ g/cm$^3$, thus requires that $z \geq 0.6$ cm. With $\lambda < z$, the condition for the ignition of a thermonuclear detonation wave is satisfied.

---

[1] For the DT reaction one must have $\rho r \geq 1$ g/cm$^2$ and $T > 10^8$ K.

The ignition energy is given by

$$E_{ign} \sim 3nkT\pi r^2 z \qquad (7)$$

For 100 fold compressed deuterium, one has $\pi r^2 = 10^{-2}$ cm$^2$, when initially it was $\pi r^2 = 10^{-1}$ cm$^2$. With $\pi r^2 = 10^{-2}$ cm$^2$, z = 0.6 cm, $kT \approx 10^{-7}$ erg ($T \sim 10^9$ K), one finds that $E_{ign} \cong 10^{16}$ erg = 1 gigajoule. This energy is provided by the $10^7$ Ampere - $10^9$ Volt proton beam lasting $10^{-7}$ seconds. The time is short enough to assure the cold compression of deuterium to high densities. For a $10^3$ fold compression, found feasible in laser fusion experiments, the ignition energy is reduced to 100 MJ.

## 5. Conclusion

One general objection against fusion in general is that it is good in producing neutrons, rather than producing heat. In nuclear fission the opposite is true. The objection is certainly valid with regard to the release of fusion energy by the DT thermonuclear reaction, the only practically possible reaction for magnetic plasma confinement fusion. It is much less valid for inertial confinement fusion, because there the ignition of a small amount of DT can ignite a larger amount of D. Still much more attractive is the ignition of a thermonuclear detonation in pure deuterium. Only with it are high gains possible. And it would eliminate the dependence on the expensive tritium. In the absence of the large magnetic field of a beam, the condition of a detonation in deuterium $\rho r > 10$ g/cm$^2$, seems only possible with a fission bomb trigger. If it should turn out to be possible to generate gigavolt potentials and with it the generation of intense ($10^7$ Ampere) GeV ion beams, one would for a magnetic field supported detonation wave along a thin deuterium rod of length z, only have to require that $\rho z \geq 10$ g/cm$^2$. With its potential to reach large yields, this would be ideal for an Orion-type pure fusion bomb propulsion system, for the transport of large payloads at high velocities within the solar system [8].